# On the RMS Anisotropy at 7° and 10° Observed in the $COBE^1$-DMR Two Year Sky Maps

A.J. Banday[2,3], K.M. Górski[2,4], L. Tenorio[5], E.L. Wright[6],
G.F. Smoot[5], C.H. Lineweaver[5], A. Kogut[7], G. Hinshaw[7],
& C.L. Bennett[8].

[2]Universities Space Research Association, NASA/Goddard Space Flight Center, Code 685.3, Greenbelt, Maryland 20771.

[3]e-mail: *banday@stars.gsfc.nasa.gov*

[4]on leave from Warsaw University Observatory, Aleje Ujazdowskie 4, 00-478 Warszawa, Poland.

[5]LBL, SSL, & CfPA, Bldg 50-351, University of California, Berkeley CA 94720.

[6]UCLA Astronomy Department, Los Angeles CA 90024-1562.

[7]Hughes STX Corporation, 4400 Forbes Blvd., Lanham MD 20706.

[8]NASA Goddard Space Flight Center, Code 685, Greenbelt MD 20771.

astro-ph/9408097  30 Aug 1994

# ABSTRACT

The frequency-independent RMS temperature fluctuations determined from the $COBE$-DMR two year sky maps are used to infer the parameter $Q_{rms-PS}$, which characterizes the normalization of power law models of primordial cosmological temperature anisotropy. In particular, a 'cross'-RMS statistic is used to determine $Q_{rms-PS}$ for a forced fit to a scale-invariant Harrison-Zel'dovich ($n = 1$) spectral model. Using a joint analysis of the 7° and 10° RMS temperature derived from both the 53 and 90 GHz sky maps, we find $Q_{rms-PS} = 17.0^{+2.5}_{-2.1}$ $\mu$K when the low quadrupole is included, and $Q_{rms-PS} = 19.4^{+2.3}_{-2.1}$ $\mu$K excluding the quadrupole. These results are consistent with the $n = 1$ fits from more sensitive methods (e.g. power spectrum, correlation function). The effect of the low quadrupole derived from the $COBE$-DMR data on the inferred $Q_{rms-PS}$ normalization is investigated. A bias to lower $Q_{rms-PS}$ is found when the quadrupole is included. The higher normalization for a forced $n = 1$ fit is then favored by the cross-RMS technique. As initially pointed out in Wright et al. (1994a) and further discussed here, analytic formulae for the RMS sky temperature fluctuations will *not* provide the correct normalization amplitude.

*Subject headings:* cosmic microwave background — cosmology: observations

## 1.  INTRODUCTION

Analysis of the first year of results from the $COBE$-DMR (Smoot et al. 1992; Bennett et al. 1992; Wright et al. 1992; Kogut et al. 1992) unambiguously demonstrated the existence of the long sought-after cosmological anisotropy in the Cosmic Microwave Background (CMB). The observed anisotropy is consistent with that predicted by models of structure formation with power law initial fluctuations of gaussian distributed amplitudes and random phases. The subsequent analysis of two years of data from the $COBE$-DMR (Bennett et al. 1994) has confirmed and refined the initial results.

In principle, the observed sky-RMS on a given angular scale is a convenient number to use in the normalization of a particular cosmological model. Previously, Wright et al. (1994a) used the sky-RMS from the first year $COBE$-DMR sky maps smoothed to an approximate FWHM of 10°, $\sigma_{sky}(10°)$, to determine the effective normalization $Q_{rms-PS}$ for the scale-invariant Harrison-Zel'dovich power spectrum, $P(k) \propto k^n$, where $n = 1$ and $k$ is the comoving wavenumber. It was also demonstrated that it is essential to account for both instrument specific details, such as the exact beam response function (rather than using a gaussian approximation, for example), and data analysis specific details, such as the subtraction of the best-fit monopole and dipole from the maps, which perturbs the inferred normalization from that derived from standard analytic formulae. An analysis based on an integral moment, such as the sky-RMS at one particular smoothing, does not have sufficient power to discriminate between different cosmological models. However, if the above details are taken into account then it can indeed provide a useful criterion for the normalization of a particular model. It should also be noted that using the observed sky-RMS values at a number of smoothing angles could be considered a poor-man's power spectrum analysis, and can be used to attempt to distinguish between models (see Smoot et al. 1994).

In this paper, we update and extend the analysis of Wright et al. (1994a) using the two year $COBE$-DMR data to infer the normalization for an $n = 1$ power law model. A cross-RMS between two maps is defined and derived either including or excluding the quadrupole. The cross-RMS is determined from the maps with no additional smoothing (which we shall refer to as the 7° smoothing, since the central lobe of the DMR beam is *approximately* described by a 7° FWHM gaussian), and after an additional 7° FWHM gaussian smoothing to 10° effective FWHM. A likelihood analysis is performed for each smoothing both individually and jointly (although the joint analysis is not used with the intention of distinguishing between different power law spectral slopes). In particular, we find that the inferred $Q_{rms-PS}$ is reasonably independent of data selection, and is consistent with those values obtained from more powerful techniques, such as power spectrum estimates (Górski 1994; Górski et al. 1994; Wright et al. 1994b) and the correlation function (Bennett et al. 1994), when restricted to $n = 1$ models.

## 2. DATA SELECTION AND SKY SIGNAL ESTIMATION

The $COBE$-DMR two year sky maps are used in the present analysis. To minimize the contribution from galactic emission, only those pixels for which the galactic latitude $\mid b \mid > 20°$ are used. Residual high latitude galactic emission and systematic error contributions are ignored (but are expected to be small, see Bennett et al. 1992, 1994). The best fit monopole and dipole and, where noted, quadrupole are also removed from the cut sky using a chi-squared technique and uniform weighting. The monopole and dipole are not physically relevant for the inference of the normalization parameter of a given anisotropy model, whilst the quadrupole is the multipole most contaminated by residual galactic emission and systematic errors in the maps. For determining the sky-RMS at 10°, the data surviving the galactic cut are smoothed by a 7° FWHM gaussian kernel with uniform weighting. GET_SKY_RMS, described in Wright et al. (1994a), uses a similar smoothing kernel, but with weighting by the number of observations per pixel. The cross-RMS, $\otimes_{RMS}$, between two maps $a$ and $b$ is then defined by

$$(\otimes_{RMS})^2 \equiv \sum_i T_i^a T_i^b\, w_i^a\, w_i^b \,/\, \sum_i w_i^a\, w_i^b, \qquad (1)$$

where the sums are over all pixels $i$ surviving the galactic cut, and $w_i$ is the weight assigned to that pixel. Smoot et al. (1994) have discussed the effects of various weighting strategies on higher order statistics in some detail. The main result is that, although weighting of order $N_i^2$ (where $N_i$ is the number of observations in a given pixel $i$) will minimize the effects of noise on the sky-RMS, such weighting will increase the effects of cosmic variance in the ensemble of simulated skies used to infer $Q_{rms-PS}$. Unit weighting is preferred here, since it is nearly optimal in the sense of minimizing the spread in the simulated RMS values, and is more easily compared to analytic techniques.

The $\otimes_{RMS}$ is a statistically unbiased estimator of the true cosmological RMS temperature anisotropy. If we write the observed temperature in a given pixel $i$, $T_i$, as the sum of a cosmic term, $t_i$, and a noise term, $\sigma_i$, then for two maps $a$ and $b$,

$$(\otimes_{RMS})^2 = \frac{1}{N_{pix}}(\sum_i t_i^2 \,+\, \sum_i t_i \sigma_i^a \,+\, \sum_i t_i \sigma_i^b \,+\, \sum_i \sigma_i^a \sigma_i^b), \qquad (2)$$

where $N_{pix}$ is the total number of pixels, and we have taken the cosmic term in both maps to be equal. The noise terms will approach zero when averaged over a large number of noise realizations. However, in any given realization of the $\otimes_{RMS}$, such as the actual data, there will be non-vanishing noise contributions. These can obviously influence the outcome of the $Q_{rms-PS}$ inference, and this issue is addressed below. Note that, under the definition of eq.(2), the $\otimes_{RMS}$ is equivalent to the estimator of excess variance defined by Wright et al. (1994a) and applied to the first year $COBE$-DMR 53 GHz maps. Its virtue is that it is more easily computed and more readily identified with the elimination of any noise contribution.

Table 1 shows the observed $\otimes_{RMS}$ values determined from the independent first and second year $COBE$-DMR sky maps at 31.5, 53 and 90 GHz. The results are consistent with the presence of sky signal in both years of data. The lack of any frequency dependence of the form $\nu^\beta$ in the $\otimes_{RMS}$ (expressed in thermodynamic temperature units) without any correction for galactic emissions is consistent with a cosmic origin for the signal. Nevertheless, the 31.5 GHz channels are the most likely to be contaminated by residual galactic emission and are also less sensitive than the 53 and 90 GHz channels. For the remainder of the analysis we only consider data at the two higher frequencies. Table 2 summarizes the $\otimes_{RMS}$ values derived from possible combinations of the 53 and 90 GHz A and B channel maps. Note that whilst the $\otimes_{RMS}$ values are generally in excellent statistical agreement, the outcome of the $Q_{rms-PS}$ analysis can be sensitive to the particular combinations selected. In the likelihood analysis below, three combinations are employed. The first, 53(A+B)$\otimes$(90A+90B), is the one used in the correlation function analysis of Bennett et al. (1994), the second, 53A$\otimes$53B, is the equivalent combination to that used in Wright et al. (1994a), and the third, (53+90)A$\otimes$(53+90)B, is one of the map combinations used in Wright et al. (1994b).

## 3. LIKELIHOOD ANALYSIS

In this *Letter* we restrict our attention to the inference of the $Q_{rms-PS}$ normalization for the $n = 1$ power spectrum model. The analytical form of the probability distribution of the $COBE$-DMR data analysis specific $\otimes_{RMS}$ statistic is unmanagable. A Monte Carlo approach was adopted to generate the $\otimes_{RMS}$ distributions for a grid of $Q_{rms-PS}$ values (with 2500 simulations used for each value of $Q_{rms-PS}$). Each simulation generates maps of the sky temperature distribution by combining a realization of Harrison-Zel'dovich sky anisotropy filtered through the $COBE$-DMR beam (Wright et al. 1994a) with noise realizations based on appropriate values of the rms per observation and observation patterns of the specific DMR channels. The effects of noise correlations are negligible (Lineweaver et al. 1994) and, hence, are not included in these simulations. The $Q_{rms-PS}$-dependent statistical means, variances and covariances of the 7° and 10° $\otimes_{RMS}$ are derived from these Monte Carlo simulations, and used to construct the gaussian approximation to the probability distribution of the $\otimes_{RMS}$. This, together with the measured $\otimes_{RMS}$ values, defines the likelihood function $\mathcal{L}(Q_{rms-PS})$.

Figure 1 shows the likelihood functions obtained from the data at 7° and 10°, and from a joint analysis of the two smoothing angles, both including and excluding the quadrupole. Table 3 summarizes the maximum likelihood values of $Q_{rms-PS}$ and the 68% and 95% confidence level intervals. These values of the rms quadrupole normalization amplitude for an $n = 1$ power law model as determined by the $\otimes_{RMS}$ technique are completely consistent with several previous analyses based on the two year 53 and 90 GHz data. Górski et al. (1994) use a power spectrum analysis and find $Q_{rms-PS} = 19.9 \pm 1.6$ $\mu$K and $20.4 \pm 1.7$ $\mu$K including or excluding the quadrupole respectively. Wright et al. (1994b) use a quadratic power spectrum estimator excluding the quadrupole to determine $Q_{rms-PS} = 19.8 \pm 2.0$ $\mu$K. In Bennett et al. (1994) the cross-correlation function is employed to infer $Q_{rms-PS} = 18.2 \pm 1.5$ $\mu$K when the quadrupole is included, and $18.6 \pm 1.6$ $\mu$K excluding the quadrupole.

It should be recalled that the first year 53 GHz 10° sky-RMS, analyzed solely with the quadrupole included, rendered the best estimate of $Q_{rms-PS} = 17.1 \pm 2.9$ $\mu$K (Wright et al. 1994a). The corresponding result from the two-year data is $17.3^{+2.5}_{-2.1}$ (see Table 3), which agrees well with the one year result (with appropriately lower errors). However, our analysis of the combined 7° and 10° $\otimes_{RMS}$ with quadrupole excluded implies a higher $Q_{rms-PS}$ normalization. Other more powerful techniques as applied to the $COBE$-DMR data for the inference of this cosmologically interesting parameter have also proven sensitive to the inclusion of the observed quadrupole, which has a relatively low value of $6 \pm 3$ $\mu$K (Bennett et al. 1994). Whether this diminuitive quadrupole amplitude has specific cosmological implications, is a result of chance cancellation of residual galactic emission with some of the cosmic quadrupole emission, or is only a manifestation of cosmic variance remains open to speculation. However, its consequences to this analysis are of interest, and have been studied with the aid of miscellaneous Monte Carlo simulations.

The particular $Q_{rms-PS} = 20$ $\mu$K model selected for scrutiny was suggested by the analysis of Górski et al. (1994). This was the most consistent, either including or excluding the quadrupole, in the inference of $Q_{rms-PS}$ for an $n = 1$ spectrum. Figure 2 shows a scatter plot of the $\otimes_{RMS}$ at 7° and 10° smoothing from 2500 simulations of noise-contaminated Harrison-Zel'dovich skies. The $\otimes_{RMS}$ data point when the quadrupole is included is $\sim 1\sigma$ deviant from the ensemble average of the simulations (Figure 2a), which is statistically satisfactory. When the quadrupole is excluded (Figure 2b), the data and simulations are in excellent agreement. So, are these to be considered mutually exclusive? A tenable argument to the contrary arises from additional simulations in which the realization-specific quadrupole is restricted to be less than 9 $\mu$K. Figure 3a demonstrates the consistency between the $\otimes_{RMS}$ data and the average over this ensemble. As seen in Figure 3b, this constraint does not affect the quadrupole-excluded case. This, then, provides a plausible explanation for the difference in $Q_{rms-PS}$ amplitudes inferred when either including, or excluding, the small observed quadrupole from the analysis. To quantify this, a sample drawn from the restricted quadrupole simulations was processed by the maximum likelihood machinery. A bias of 2 - 3 $\mu$K was observed to lower $Q_{rms-PS}$ (depending on which smoothing angles were involved) when the quadrupole was included. The bias corrected maximum likelihood $Q_{rms-PS}$ normalizations are 20.1, 19.5 and 19.7 $\mu$K for the 7°, 10° and joint analysis respectively. No bias was observed when the test data were analyzed excluding the quadrupole. Thus, we can conclude that if the bias introduced by the low observed quadrupole is accounted for, then the normalizations inferred with the quadrupole either included or excluded are consistent in the context of a forced $n = 1$ fit. A $Q_{rms-PS}$ value of 19.4 $\mu$K is the most appropriate for the $\otimes_{RMS}$ analysis of an $n = 1$ spectrum, after accounting for the bias due to the low quadrupole.

A further important aspect of the analysis demonstrated in Figure 2 is the comparison with analytic calculations of the RMS. The analytic formula for the sky-variance, with a gaussian approximation to the beam filter function, specified for an $n = 1$ model, is

$$\langle (\Delta T)^2 \rangle \; = \; 1.2 \sum_\ell Q^2_{rms-PS} \, \frac{(2\ell+1)}{\ell(\ell+1)} \, e^{-\ell(\ell+1)\sigma_b^2} \qquad (3)$$

where $\ell$ is the spherical harmonic order and $\sigma_b$ is the gaussian beam dispersion. Here, $\sigma_b$ = 3° corresponding to the approximate 7° FWHM $COBE$-DMR beam. The sum over $\ell$ is taken in the range [2,40] or [3,40], which is sufficient to determine the $\otimes_{RMS}$ either including or excluding the quadrupole respectively. This formula overestimates the strength of the fluctuations, since the actual $COBE$-DMR beam filter function drops in amplitude in $\ell$-space more rapidly than the gaussian approximation (see Wright et al. 1994a). Even when the correct beam and pixelization weights are included, there remains a disagreement between the analytic calculations and the simulations that explicitly account for the monopole, dipole and, if required, quadrupole subtraction from the galaxy-cut data. Since these multipole estimates are made on incomplete sky coverage, there is some aliasing of higher order power into the fitted and removed low order amplitudes, thus the RMS fluctuation amplitudes are suppressed. The combined beam filter and multipole subtraction effects are of order 5-10% when the quadrupole is included (in agreement with Wright et al. 1994a), and 10-15% in the no quadrupole case.

## 4. DISCUSSION

Some aspects of the analysis related to a proper assessment of the results include: 1) choice of input data values, 2) noise model uncertainties, 3) biases in the parameter inference and 4) relation to other (power law) cosmological models.

We have only used three of the cross-combinations possible with the 53 and 90 GHz data in this analysis, as motivated by previous work. Inspection of the observed $\otimes_{RMS}$ values in Table 2, together with the errors on the inferred $Q_{rms-PS}$ values, suggests that all of the combinations should be consistent with $Q_{rms-PS} \sim 19$ $\mu$K with the possible exception of the 90A$\otimes$90B 7° RMS. However, this is most likely just an anomaly due to noise: in 2500 simulations generated with $Q_{rms-PS} = 19$ $\mu$K, this particular combination yielded a zero $\otimes_{RMS}$ in 7.0% of the simulations including the quadrupole, and in 12.2% of those excluding the quadrupole. Further, although the 7° $\otimes_{RMS}$ prefers a $Q_{rms-PS}$ normalization of zero, it is still consistent with the 19.4 $\mu$K normalization determined previously at the $\sim 2\sigma$ level. The 10° and joint analyses of the 90 GHz data, both when the quadrupole is included or excluded, are consistent with this normalization to $\sim 1\sigma$. An analysis including both the 53 and 90 GHz data is to be preferred due to its higher sensitivity.

The noise RMS per observation, $\sigma_{obs}$, is known to an accuracy of $\sim 1\%$. Simulations performed with $\sigma_{obs}$ adjusted by such an amount demonstrate that the maximum likelihood values for $Q_{rms-PS}$ are shifted by $\sim 0.1$ $\mu$K, an insignificant amount compared to the error in the inferred quadrupole normalization.

The particular observed $\otimes_{RMS}$ value from the 53 and 90 GHz data is noise contaminated (eq.2). The important issue is to determine if this results in a biased inference of $Q_{rms-PS}$. A sub-ensemble of the simulations was selected and used as test input data. No statistically significant bias was observed in the simulated sample-averaged estimates of $Q_{rms-PS}$ either including or excluding the quadrupole.

This analysis has been specific to a forced n = 1 spectral fit. The normalization of other models of cosmological anisotropy should proceed either by a detailed reworking of the above, or by using more powerful techniques that are sensitive to both $Q_{rms-PS}$ and $n$, such as the power spectrum method (Górski et al. 1994; Wright et al. 1994b) or the 2-point correlation function (Bennett et al. 1994). In fact, the analysis of Górski et al. (1994) has rendered a useful power spectrum independent normalization, $a_9 \sim 8$ $\mu$K, for power law spectral models, whilst the two-point correlation function technique described in Bennett et al. (1994) finds $a_7 \sim 9.5$ $\mu$K.

An exact calculation for the power spectrum of CMB anisotropy that includes all the potential, velocity and adiabatic effects on the last scattering surface in the context of a primordial $n = 1$ spectrum renders an effective spectral slope slightly steeper than $n = 1$ over the angular scales probed by the $COBE$-DMR instrument. However, since such an exactly computed power spectrum is not sufficiently described by a simple power law, we have implemented an analysis as above using multipole coefficients, kindly provided by Radek Stompor, for a specific $h = 0.5$, $\Omega_b = 0.03$, CDM model. The $Q_{rms-PS}$ normalization inferred from the joint 7° and 10° analysis of the 53 and 90 GHz data is $16.7^{+2.4}_{-1.9}$ $\mu$K including the quadrupole, and $18.9^{+2.2}_{-1.9}$ $\mu$K when the quadrupole is excluded. As discussed previously, the latter higher value is preferred in the present analysis for the low quadrupole case at hand.

In summary, we have used the $\otimes_{RMS}$ statistic derived from the $COBE$-DMR two year 53 and 90 GHz maps to infer a normalization $Q_{rms-PS} \sim 19$ $\mu$K for the amplitude of primordial inhomogeneity in the context of a Harrison-Zel'dovich $n = 1$ model. We stress that simple analytic models that do not include a correct description of the $COBE$-DMR beam or the monopole, dipole and, if appropriate, quadrupole subtraction will underestimate the amplitude of $Q_{rms-PS}$ when normalizing to the observed sky-RMS. The low observed quadrupole amplitude affords a reasonable explanation for the difference in inferred $Q_{rms-PS}$ amplitudes when either including or excluding the quadrupole.

We gratefully acknowledge the efforts of those contributing to the $COBE$ DMR. $COBE$ is supported by the Office of Space Sciences of NASA Headquarters. We thank Radek Stompor for providing us with the CDM model CMB anisotropy power spectrum coefficients computed using his Boltzmann code.

Table 1: Observed $\otimes_{RMS}$ in thermodynamic temperature derived from the first and second year 31.5, 53 and 90 GHz $COBE$-DMR data. The errors are determined from a large number of noise simulations for a fixed CMB sky realization. The frequency independence of the data, expressed as $\otimes_{RMS}(\nu) \propto \nu^\beta$ is demonstrated.

| 1st year ⊗ 2nd year | including Quadrupole | | excluding Quadrupole | |
|---|---|---|---|---|
| $\otimes_{RMS}$ combination | 7° ($\mu$K) | 10° ($\mu$K) | 7° ($\mu$K) | 10° ($\mu$K) |
| 31(A+B) | $39.7^{+27.3}_{-37.0}$ | $38.2^{+7.6}_{-7.2}$ | $34.8^{+29.0}_{-34.8}$ | $33.2^{+7.9}_{-7.5}$ |
| 53(A+B) | $35.3^{+4.8}_{-4.7}$ | $32.7^{+1.8}_{-1.8}$ | $33.5^{+5.5}_{-5.1}$ | $30.9^{+1.9}_{-1.9}$ |
| 90(A+B) | $25.9^{+12.4}_{-11.3}$ | $32.7^{+3.4}_{-3.5}$ | $25.1^{+14.1}_{-13.9}$ | $32.2^{+3.5}_{-3.5}$ |
| Frequency dependence: $\beta$ | $-0.45 \pm 0.8$ | $-0.1 \pm 0.2$ | $-0.3 \pm 0.9$ | $0.1 \pm 0.2$ |

Table 2: Observed $\otimes_{RMS}$ values derived from possible 53 and 90 GHz combinations. Those combinations used in the $Q_{rms-PS}$ analysis are denoted by '†'.

| $\otimes_{RMS}$ combination | including Quadrupole | | excluding Quadrupole | |
|---|---|---|---|---|
| | 7° ($\mu$K) | 10° ($\mu$K) | 7° ($\mu$K) | 10° ($\mu$K) |
| 53A⊗53B† | $44.5^{+4.8}_{-4.7}$ | $32.4^{+1.8}_{-1.8}$ | $43.1^{+5.4}_{-5.2}$ | $30.6^{+1.9}_{-1.9}$ |
| 90A⊗90B | $0.0^{+12.0}_{-10.9}$ | $25.7^{+3.4}_{-3.4}$ | $0.0^{+13.7}_{-13.4}$ | $24.7^{+3.5}_{-3.5}$ |
| 53A⊗90A | $28.4^{+8.1}_{-7.7}$ | $30.9^{+2.7}_{-2.9}$ | $27.7^{+9.2}_{-8.6}$ | $30.3^{+2.8}_{-2.8}$ |
| 53B⊗90B | $32.3^{+7.0}_{-6.7}$ | $29.9^{+2.3}_{-2.4}$ | $30.2^{+8.2}_{-7.6}$ | $27.7^{+2.4}_{-2.5}$ |
| 53A⊗90B | $45.2^{+6.1}_{-5.9}$ | $32.4^{+2.2}_{-2.2}$ | $44.0^{+6.9}_{-6.6}$ | $30.9^{+2.2}_{-2.3}$ |
| 53B⊗90A | $34.6^{+9.4}_{-8.7}$ | $31.7^{+2.9}_{-3.0}$ | $33.7^{+10.8}_{-10.0}$ | $30.8^{+3.0}_{-3.0}$ |
| 53(A+B)⊗90A | $31.6^{+6.3}_{-6.0}$ | $31.3^{+2.5}_{-2.6}$ | $30.8^{+7.1}_{-6.6}$ | $30.6^{+2.5}_{-2.5}$ |
| 53(A+B)⊗90B | $39.3^{+4.8}_{-4.7}$ | $31.2^{+1.9}_{-2.0}$ | $37.7^{+5.3}_{-5.1}$ | $29.3^{+2.0}_{-2.0}$ |
| 53A⊗90(A+B) | $37.7^{+5.1}_{-4.9}$ | $31.7^{+1.9}_{-1.9}$ | $36.8^{+5.7}_{-5.5}$ | $30.6^{+2.0}_{-2.0}$ |
| 53B⊗90(A+B) | $33.5^{+5.8}_{-5.7}$ | $30.8^{+2.1}_{-2.1}$ | $31.9^{+6.7}_{-6.3}$ | $29.3^{+2.1}_{-2.1}$ |
| 53(A+B)⊗90(A+B)† | $35.7^{+4.0}_{-3.9}$ | $31.2^{+1.6}_{-1.7}$ | $34.4^{+4.4}_{-4.3}$ | $29.9^{+1.7}_{-1.7}$ |
| (53+90)A⊗(53+90)B† | $35.6^{+4.3}_{-4.2}$ | $30.7^{+1.7}_{-1.7}$ | $34.4^{+4.9}_{-4.6}$ | $29.4^{+1.7}_{-1.7}$ |
| (53A+90B)⊗(53B+90A) | $30.3^{+4.3}_{-4.3}$ | $29.9^{+1.7}_{-1.7}$ | $28.8^{+4.8}_{-4.6}$ | $28.5^{+1.7}_{-1.7}$ |

Table 3: Derived parameters from the likelihood ($\mathcal{L}$) analysis, assuming an $n = 1$ spectrum.

| 53(A+B)⊗90(A+B) | including Quadrupole | | | excluding Quadrupole | | |
|---|---|---|---|---|---|---|
| $Q_{rms-PS}$ ($\mu$K) | 7° | 10° | joint | 7° | 10° | joint |
| Maximum $\mathcal{L}$ value | 18.0 | 16.6 | 17.0 | 20.2 | 19.0 | 19.4 |
| 68% c.l. interval | [14.7, 21.2] | [14.7, 19.1] | [14.9, 19.5] | [16.6, 23.5] | [17.1, 21.3] | [17.3, 21.7] |
| 95% c.l. interval | [9.8, 24.5] | [12.9, 22.2] | [13.1, 22.9] | [10.9, 26.5] | [15.2, 24.1] | [15.5, 24.6] |

| 53A⊗53B | including Quadrupole | | | excluding Quadrupole | | |
|---|---|---|---|---|---|---|
| $Q_{rms-PS}$ ($\mu$K) | 7° | 10° | joint | 7° | 10° | joint |
| Maximum $\mathcal{L}$ value | 22.3 | 17.3 | 16.7 | 25.0 | 19.4 | 19.1 |
| 68% c.l. interval | [18.7, 25.8] | [15.2, 19.8] | [14.7, 19.0] | [21.3, 28.6] | [17.4, 21.9] | [17.0, 21.4] |
| 95% c.l. interval | [13.8, 29.9] | [13.3, 23.1] | [13.0, 22.3] | [15.5, 32.4] | [15.4, 24.8] | [15.1, 24.3] |

| (53+90)A⊗(53+90)B | including Quadrupole | | | excluding Quadrupole | | |
|---|---|---|---|---|---|---|
| $Q_{rms-PS}$ ($\mu$K) | 7° | 10° | joint | 7° | 10° | joint |
| Maximum $\mathcal{L}$ value | 18.1 | 16.6 | 17.0 | 20.4 | 19.0 | 19.4 |
| 68% c.l. interval | [14.5, 21.4] | [14.7, 19.1] | [14.9, 19.6] | [16.4, 23.8] | [17.0, 21.3] | [17.3, 21.7] |
| 95% c.l. interval | [9.3, 24.7] | [12.9, 22.3] | [13.1, 23.0] | [10.3, 26.8] | [15.2, 24.2] | [15.4, 24.7] |

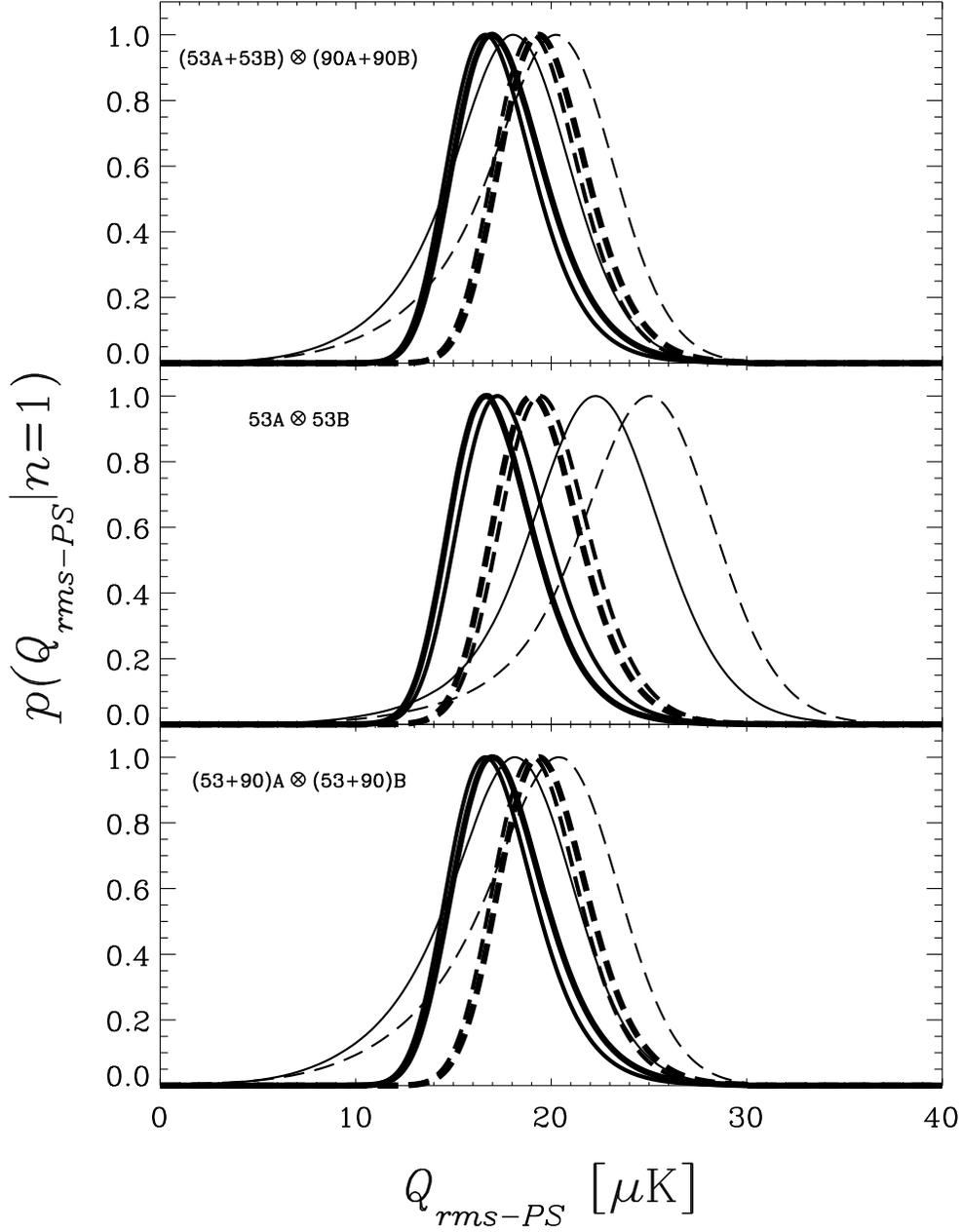

Fig. 1.— Likelihood curves for $Q_{rms-PS}$ derived from the $\otimes_{RMS}$ assuming $n = 1$. Thin lines represent the analysis based on the 7° RMS, thicker lines from the 10° RMS, and thickest lines from the joint analysis. Solid curves include the quadrupole, the dashed curves are for the no quadrupole case. Top: 53(A+B)⊗90(A+B), center: 53A⊗53B, bottom: (53+90)A⊗(53+90)B.

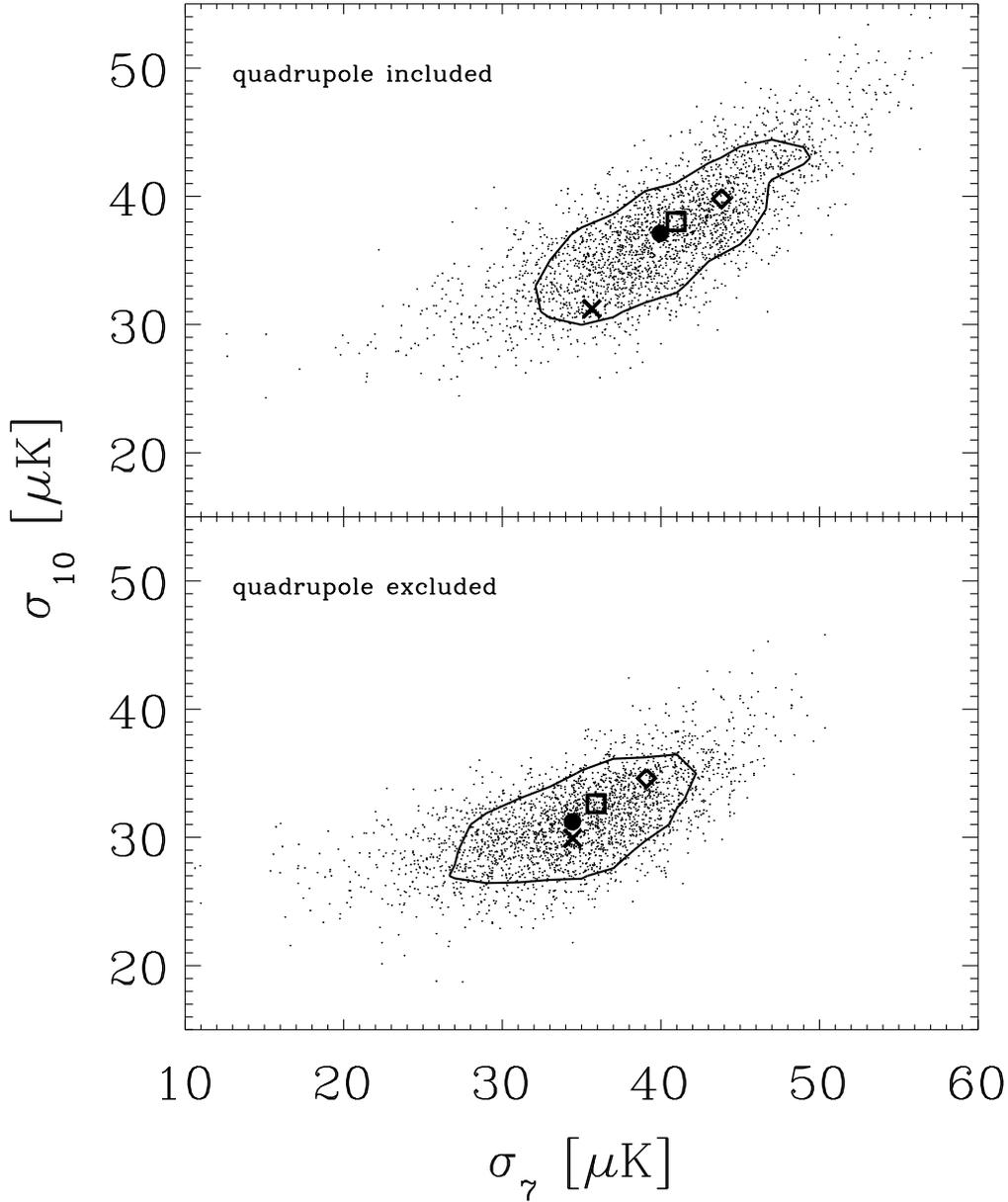

Fig. 2.— Scatter plots for 2500 simulations of the 7° and 10° $\otimes_{RMS}$ for $Q_{rms-PS} = 20$ μK and $n = 1$ with the noise properties of the 53 and 90 GHz (A+B) maps. The cross represents the 53(A+B)⊗90(A+B) data, the filled circle is the mean from the simulated ensemble, the square is the analytic prediction using the $COBE$-DMR beam and pixelization smoothing, and the diamond is the analytic prediction using a gaussian approximation to the $COBE$-DMR beam and no pixelization smoothing. Also shown is the 1$\sigma$ density contour derived from the simulations. Top: quadrupole included, bottom: quadrupole excluded.

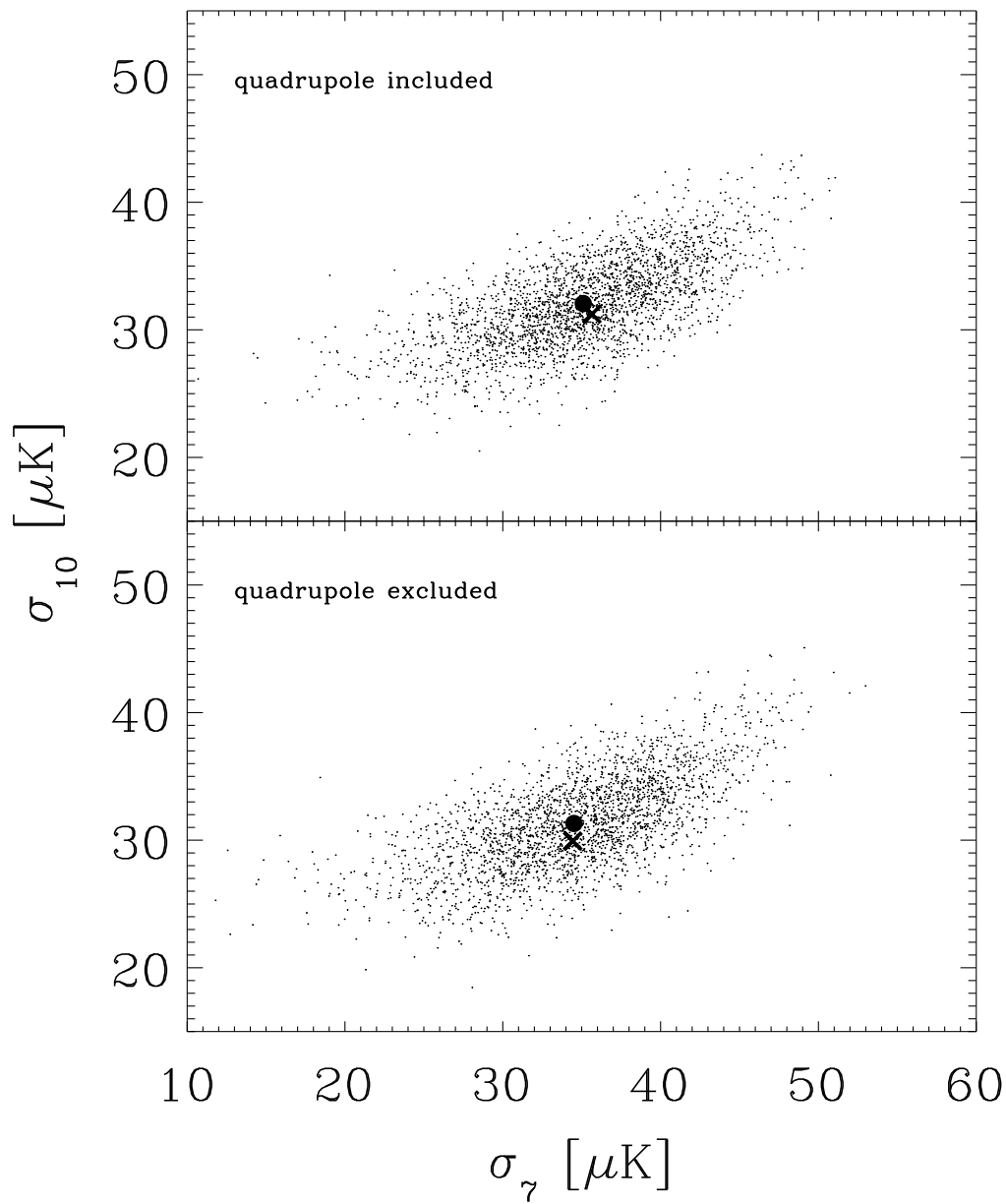

Fig. 3.— As Figure 2, but with the ensemble comprising 2500 realizations with a realization-specific quadrupole less than 9 $\mu$K. The cross represents the 53(A+B)⊗90(A+B) data, and the filled circle is the mean from the simulated ensemble. Top: quadrupole included, bottom: quadrupole excluded.